\documentclass[a4paper,11pt]{article}%
\usepackage{geometry}
\usepackage{amsmath}
\usepackage{amsfonts}
\usepackage{amssymb}
\usepackage{graphicx}
\usepackage{indentfirst}
\usepackage[small,bf]{caption}
\usepackage{slashed}
\providecommand{\U}[1]{\protect\rule{.1in}{.1in}}

\begin{document}

\date{}
\title{\textbf{Dimension two condensates in the Gribov-Zwanziger theory in the Coulomb gauge}}
\author{\textbf{M.~S.~Guimaraes}\thanks{msguimaraes@uerj.br}\,\,,
\textbf{B.~W.~Mintz}\thanks{bruno.mintz.uerj@gmail.com}\,\,,
\textbf{S.~P.~Sorella}\thanks{silvio.sorella@gmail.com}\,\,\,,\\[2mm]
{\small \textnormal{  \it Departamento de F\'{\i }sica Te\'{o}rica, Instituto de F\'{\i }sica, UERJ - Universidade do Estado do Rio de Janeiro,}}
 \\ \small \textnormal{ \it Rua S\~{a}o Francisco Xavier 524, 20550-013 Maracan\~{a}, Rio de Janeiro, Brasil}\normalsize}
\maketitle

\begin{abstract}
We investigate the dimension two condensate $\langle {\bar \phi}^{ab}_i \phi^{ab}_i -{\bar \omega}^{ab}_i \omega^{ab}_i \rangle$  
within the Gribov-Zwanziger approach to Euclidean Yang-Mills theories in the Coulomb gauge,  in both  $3$ and $4$ dimensions. 
An explicit calculation shows that, at the first order, the condensate 
$\langle {\bar \phi}^{ab}_i \phi^{ab}_i -{\bar \omega}^{ab}_i \omega^{ab}_i \rangle$ is plagued by a non-integrable IR divergence 
in $3D$, while in $4D$ it exhibits a logarithmic UV divergence, being  proportional to the Gribov parameter $\gamma^2$. These results 
indicate that in 3D the  transverse spatial Coulomb gluon two-point correlation function exhibits a scaling behaviour, in agreement with Gribov's 
expression. In 4D, however, they suggest that, next to the scaling behaviour, a decoupling solution might emerge too.  

\end{abstract}

\section{Introduction} 
Th Coulomb gauge, $\partial_i A^a_i=0$, $i=1,...., D-1$, is largely employed in analytic 
\cite{Cucchieri:1996ja,Cucchieri:2000hv,Zwanziger:2002sh,Zwanziger:2003de,Greensite:2004ke,Zwanziger:2004np,Greensite:2004ur,Zwanziger:2006sc,Schleifenbaum:2006bq,Watson:2006yq,Epple:2006hv,Watson:2007vc,Campagnari:2009wj,Burgio:2009xp,Watson:2010cn,Leder:2011yc,Weber:2011zzd,Szczepaniak:2011bs,Watson:2011kv,Campagnari:2011bk,Watson:2012ht,Huber:2014isa} as well as in lattice numerical investigations \cite{Cucchieri:2000gu,Cucchieri:2000kw,Burgio:2008jr,Nakagawa:2009zf,Burgio:2012bk} of non-perturbative aspects of  Euclidean yang-Mills theories. An impressive number of results are nowadays available in this gauge, providing a consistent scenario for confinement. 
\\\\In this letter we focus on aspects of the Gribov-Zwanziger formulation of the Coulomb gauge \cite{Gribov:1977wm,Zwanziger:2002sh,Zwanziger:2003de,Zwanziger:2004np,Zwanziger:2006sc}, 
which implements the restriction of the domain of integration in the 
functional Euclidean integral to the Gribov Region $\Omega$, defined as the set of all field configurations fulfilling the Coulomb condition and for which the Faddeev-Popov operator, 
${\cal M}^{ab}= -\delta^{ab}\partial_i \partial_i -g f^{acb}A^c_i\partial_i$, is strictly positive, namely 
\begin{equation}
\Omega = \{ A^a_i \;, \;\; \partial_iA^a_i =0\;, \;\;  {\cal M}^{ab} >0 \;\; \}  \;.  \label{gr}
\end{equation}
The restriction to the region $\Omega$ accounts for the existence of the Gribov copies, which affect the Coulomb gauge. The so called Gribov-Zwanziger action 
\cite{Gribov:1977wm,Zwanziger:2002sh,Zwanziger:2003de,Zwanziger:2004np,Zwanziger:2006sc} is the final result of the restriction to the region $\Omega$. Besides 
the Coulomb gauge, the restriction to the Gribov region has been effectively implemented in the Landau  \cite{Gribov:1977wm,Zwanziger:1988jt,Zwanziger:1989mf} 
and maximal Abelian gauges \cite{Capri:2006vv,Capri:2006cz}, where the corresponding Gribov-Zwanziger actions have been worked out. A feature of the Gribov-Zwanziger 
set up in the Landau and maximal Abelian gauges is that the two-point gluon correlation function is strongly suppressed in the infrared region in momentum space $k$, 
attaining a vanishing value when $k=0$. This kind of behaviour is usually referred to as the scaling solution, also observed in the study of the Dyson-Schwinger  
equations in these gauges, see  \cite{Alkofer:2000wg,Huber:2009wh}. Such a behaviour is also found for the spatial gluon correlation function in the Coulomb gauge. 
\\\\Nevertheless, next to the scaling solution, another type of solution, called decoupling solution, was found in both Landau 
\cite{Cornwall:1981zr,Aguilar:2004sw,Dudal:2007cw,Aguilar:2008xm,Dudal:2008sp,Dudal:2008rm,Fischer:2008uz} and maximal Abelian gauges  
\cite{Capri:2008ak,Capri:2008vk,Capri:2010an}, in both $3$ and $4$ dimensions.  Similarly to the scaling solution, the decoupling 
solution is also strongly suppressed in the infrared region, displaying a violation of positivity. However, it does not attain a vanishing  value at $k=0$. 
\\\\Within the Gribov-Zwanziger approach, the decoupling solution arises as the consequence of the existence of dimension two condensates 
\cite{Dudal:2007cw,Dudal:2008sp,Dudal:2008rm,Capri:2008ak,Capri:2008vk,Capri:2010an}. The resulting action accounting for the inclusion of 
these condensates is known as the Refined-Gribov-Zwanziger action \cite{Dudal:2007cw,Dudal:2008sp,Dudal:2008rm,Capri:2008ak,Capri:2008vk,Capri:2010an}. 
\\\\In the present work, we investigate the condensate $\langle {\bar \phi}^{ab}_i \phi^{ab}_i -{\bar \omega}^{ab}_i \omega^{ab}_i \rangle$  in Coulomb gauge, 
in both $3$ and $4$ dimensions. In $3D$ we find, by an explicit first order calculation, that the integral defining 
$\langle {\bar \phi}^{ab}_i \phi^{ab}_i -{\bar \omega}^{ab}_i \omega^{ab}_i \rangle$ exhibits a non-integrable IR divergence, showing that the condensate cannot 
be safely introduced in $3D$. As a consequence, in $3D$, the transverse equal-time gluon propagator exhibits the scaling type behaviour given by Gribov's expression. 
\\\\Moreover, in $4D$, we find that the condensate is safe in the IR, being plagued by a mild UV logarithmic divergence, precisely as the gap equation defining the 
Gribov parameter. This suggests that, apart from a UV proper renormalization,  a non-vanishing dimension two condensate 
$\langle {\bar \phi}^{ab}_i \phi^{ab}_i -{\bar \omega}^{ab}_i \omega^{ab}_i \rangle$ might show up also in the Coulomb gauge. 
As a consequence, the spatial transverse two-point gluon correlation function exhibits a decoupling solution, next to the well known scaling one. To some extent, 
this suggests a kind of common feature of the Landau, Coulomb and maximal Abelian gauge in $4D$ Euclidean space-time. \\\\The present letter is organised as follows. 
In Sect.2 we give a short overview of the Gribov-Zwanziger action in the Coulomb gauge. Sect.3 is devoted to the evaluation of the condensate  
$\langle {\bar \phi}^{ab}_i \phi^{ab}_i -{\bar \omega}^{ab}_i \omega^{ab}_i \rangle$ at the first order, in both $3D$ and $4D$. 

\section{The Gribov-Zwanziger action in the Coulomb gauge} 
The Gribov-Zwanziger action implementing the restriction to the Gribov region $\Omega$, eq.\eqref{gr}, in the Coulomb gauge, $\partial_i A^a_i=0$, $i=1,...,(D-1)$, reads \cite{Gribov:1977wm,Zwanziger:2002sh,Zwanziger:2003de,Zwanziger:2004np,Zwanziger:2006sc}
\begin{eqnarray}
S_{GZ} & = & \int d^Dx  \left( \frac{1}{4} F^{a}_{\mu\nu} F^{a}_{\mu\nu} + b^a \partial_i A^a_i + {\bar c}^a \partial_i D_{i}^{ab} c^b \right)  \nonumber \\
&+&  \int d^Dx \left( - {\bar \phi}^{ab}_{\mu} \partial_i D_{i}^{ab} \phi^{cb}_\mu  + {\bar \omega}^{ab}_{\mu} \partial_i D_{i}^{ab} \omega^{cb}_\mu  
- g f^{acm} (\partial_i {\bar \omega}^{ab}_{\mu})  (D_{i}^{cp} c^p)  \phi^{mb}_\mu \right)   \nonumber \\
&+& \int d^Dx \left( \gamma^2 g f^{abc} A^a_i (\phi^{bc}_i - {\bar \phi}^{bc}_i )  - (D-1) (N^2-1) \gamma^4 \right) \;. \label{gza} 
\end{eqnarray}
The field $b^a$ is the Lagrange multiplier enforcing the Coulomb condition, $\partial_i A^a_i=0$, while the fields $({\bar c}^a, c^a)$ are the Faddeev-Popov ghosts. 
The fields $({\bar \phi}^{ab}_{\mu}, { \phi}^{ab}_{\mu})$, $\mu=1,...,D$, are a set of commuting fields while $({\bar \omega}^{ab}_{\mu}, { \omega}^{ab}_{\mu})$ are 
anti-commuting. These fields are introduced in order to implement the restriction to the region $\Omega$ through a local action, eq.\eqref{gza}. Finally, the parameter 
$\gamma^2$ appearing in expression \eqref{gza} is the Gribov parameter. It has the dimension of mass  squared  and has a dynamical origin, being determined 
in a self consistent way through the gap equation 
\begin{equation}
\frac{\partial {\cal E}_v}{\partial \gamma^2}= 0 \;, \label{ev}
\end{equation}
where ${\cal E}_v$ is the vacuum energy, namely 
\begin{equation}
e^{-V {\cal E}_v} = \int [{\cal D}\Phi] \; e^{-S_{GZ}} \;. \label{z}
\end{equation}
To the first order 
\begin{equation}
{\cal E}_v = -(D-1) (N^2-1) \gamma^4 + \frac{(D-2)}{2} (N^2-1) \int \frac{d^Dk}{(2\pi)^D} \; \log \left(k^2_D +{\vec k}^2 + \frac{2Ng^2\gamma^4}{{\vec k}^2} \right)  \; + \; \;\;{\rm terms \; independent\; from}\; \;  \gamma^2 \;,  
\end{equation}
so that the gap equation, eq.\eqref{ev}, takes the form 
\begin{equation}
\int \frac{d^Dk}{(2\pi)^D} \frac{1}{k^2_D\; {\vec k}^2 + {\vec k}^4 +{2Ng^2\gamma^4} }= \frac{(D-1)}{(D-2)} \frac{1}{2Ng^2} \;. \label{gp1}
\end{equation}
From the Gribov-Zwanziger action, it follows that the tree level two-point gluon spatial correlation function is given by 
\begin{equation}
\langle A^a_i(k) A^b_j(-k) \rangle = \frac{\delta^{ab}}{ k^2_D +{\vec k}^2 + \frac{2Ng^2\gamma^4}{{\vec  k}^2}} \left( \delta_{ij} -\frac{k_i k_j} {{\overrightarrow k}^2} \right) \;, \label{gzgluon}
\end{equation}
leading to an equal-time transverse form factor \cite{Burgio:2012bk} 
\begin{equation}
D^{tr}_{GZ} ({\overrightarrow k}) = \frac{|{\overrightarrow k}|}{\sqrt{{\overrightarrow k}^4 + 2Ng^2\gamma^4}}  \;, \label{trgz}
\end{equation}
which exhibits the scaling behaviour, $D^{tr}_{GZ} (0)=0$. Before going any further, it is worth reminding briefly how expression \eqref{gzgluon} is derived. A simple calculation shows in fact that the quadratic part of the Gribov-Zwanziger action in the gluon sector takes the following form
\begin{equation}
S_{GZ}^{\rm quadr-gluon}   = \int d^Dx \;\left( \frac{1}{2} A^a_D (-{\vec{\partial}}^2\;) A^a_D + \frac{1}{2} A^a_i \left(-\partial_D^2 -{\vec{\partial}}^2+ \frac{2Ng^2\gamma^4}{-{\vec{\partial}}^2}\right) A^a_i +b^a \partial_i A^a_i  \right) \;, \label{quadr}
\end{equation}
from which one immediately derives the tree level expression for the transverse spatial gluon propagator given in eq.\eqref{gzgluon}. Moreover, one has to observe that, unlike expression \eqref{gzgluon}, the tree level temporal correlator $\langle A^a_D(k) A_D^b(-k)\rangle$ does not display an energy resolution, due to the well known existence of residual temporal gauge transformations which affect the Coulomb condition. The resolution of the energy behaviour of this correlation function is a quite delicate point which requires a detailed mastering of the renormalization procedure of the Coulomb gauge. Here, we remind the reader to the large literature existing on this subject, both in the continuum as well as in the lattice formulation  \cite{Cucchieri:1996ja,Cucchieri:2000hv,Zwanziger:2002sh,Zwanziger:2003de,Greensite:2004ke,Zwanziger:2004np,Greensite:2004ur,Zwanziger:2006sc,Schleifenbaum:2006bq,Watson:2006yq,Epple:2006hv,Watson:2007vc,Campagnari:2009wj,Burgio:2009xp,Watson:2010cn,Leder:2011yc,Weber:2011zzd,Szczepaniak:2011bs,Watson:2011kv,Campagnari:2011bk,Watson:2012ht,Huber:2014isa,Cucchieri:2000gu,Cucchieri:2000kw,Burgio:2008jr,Nakagawa:2009zf,Burgio:2012bk}.

\section{The dimension two condensate $\langle {\bar \phi}^{ab}_i \phi^{ab}_i -{\bar \omega}^{ab}_i \omega^{ab}_i \rangle$}
In order to evaluate the condensate $\langle {\bar \phi}^{ab}_i \phi^{ab}_i -{\bar \omega}^{ab}_i \omega^{ab}_i \rangle$, we couple the operator $ \left({\bar \phi}^{ab}_i(x) \phi^{ab}_i(x) -{\bar \omega}^{ab}_i(x) \omega^{ab}_i(x) \right)$ to a constant source $J$, {\it i.e.}
\begin{equation}
S_{GZ} \rightarrow S_{GZ} + \int d^Dx  J \left({\bar \phi}^{ab}_i(x) \phi^{ab}_i(x) -{\bar \omega}^{ab}_i(x) \omega^{ab}_i(x) \right)   \;, \label{jact} 
\end{equation}
and we evaluate 
\begin{equation}
e^{-V {\cal E}_v(J) } = \int [{\cal D}\Phi] \; e^{-\left(S_{GZ}+ \int d^Dx  J \left({\bar \phi}^{ab}_i(x) \phi^{ab}_i(x) -{\bar \omega}^{ab}_i(x) \omega^{ab}_i(x) \right)\right)} \;. \label{zj}
\end{equation}
The condensate is thus obtained by differentiating ${\cal E}_v(J)$ with respect to $J$ and setting $J=0$ at the end, namely 
\begin{equation}
\frac{\partial {\cal E}_v(J)}{\partial J}\Bigl|_{J= 0} = \frac{ \int [{\cal D}\Phi] \; \left({\bar \phi}^{ab}_i(x) \phi^{ab}_i(x) -{\bar \omega}^{ab}_i(x) \omega^{ab}_i(x) \right) \;e^{-S_{GZ}}}{\int [{\cal D}\Phi]  \;e^{-S_{GZ}}} = \langle {\bar \phi}^{ab}_i \phi^{ab}_i -{\bar \omega}^{ab}_i \omega^{ab}_i \rangle \;. \label{cond}
\end{equation}
At the first order, we get 
\begin{equation}
{\cal E}_v(J) = \frac{(D-2)}{2} (N^2-1) \int \frac{d^Dk}{(2\pi)^D} \; \log \left(k^2_D +{\vec k}^2 + \frac{2Ng^2\gamma^4}{{\vec k}^2+J} \right)  \; + \; \;\;{\rm terms \; independent\; from}\; \;  J \;.   \label{[j1}
\end{equation}
Taking the derivative with respect to $J$ and setting $J=0$, it turns out that 
\begin{equation}
 \langle {\bar \phi}^{ab}_i \phi^{ab}_i -{\bar \omega}^{ab}_i \omega^{ab}_i \rangle = - Ng^2\gamma^4 (N^2-1)    (D-2)
 \int \frac{d^Dk}{(2\pi)^D} \frac{1}{k^2_D\;{\vec k}^2 + {\vec k}^4 + {2Ng^2\gamma^4}} \;\frac{1}{ {\vec k}^2}   \;. \label{j2} 
\end{equation}
\subsection{The $3D$ case} 
Let us start by considering first expression \eqref{j2} in $3D$. Here, we have 
\begin{equation}
 \langle {\bar \phi}^{ab}_i \phi^{ab}_i -{\bar \omega}^{ab}_i \omega^{ab}_i \rangle = - Ng^2\gamma^4 (N^2-1)  
 \int \frac{dk_3\; d^2{\vec k} }{(2\pi)^3} \frac{1}{k^2_3\;{\vec k}^2 + {\vec k}^4 + {2Ng^2\gamma^4}} \;\frac{1}{ {\vec k}^2}   \;,  \label{cd31} 
\end{equation}
while for the gap equation 
\begin{equation}
\int \frac{dk_3 \;d^2{\vec k}}{(2\pi)^3} \frac{1}{k^2_3\; {\vec k}^2 + {\vec k}^4 +{2Ng^2\gamma^4} }=  \frac{1}{Ng^2} \;. \label{gp31}
\end{equation}
We see that, while the integral defining the gap equation, eq.\eqref{gp31}, is convergent in both IR and UV regions, the expression for the condensate, eq.\eqref{cd31}, exhibits a non-integrable singularity in the IR, due to the presence of the term  $\frac{1}{ {\vec k}^2}$ which is non-integrable in two-dimensional space. This indicates that the condensate $\langle {\bar \phi}^{ab}_i \phi^{ab}_i -{\bar \omega}^{ab}_i \omega^{ab}_i \rangle$ cannot be introduced in $3D$, due to the existence of infrared divergences. A similar phenomenon occurs in the Landau gauge in $2D$, where the analogous condensate cannot be introduced due to infrared divergences \cite{Dudal:2008xd}.  
\subsection{The $4D$ case} 
Let us turn now to the $4D$ case, where for the condensate and for the gap equation we get 
\begin{equation}
  \langle {\bar \phi}^{ab}_i \phi^{ab}_i -{\bar \omega}^{ab}_i \omega^{ab}_i \rangle = - 2Ng^2\gamma^4 (N^2-1)   
 \int \frac{dk_4 \;d^3{\vec k}}{(2\pi)^4} \frac{1}{k^2_4\;{\vec k}^2 + {\vec k}^4 + {2Ng^2\gamma^4}} \;\frac{1}{ {\vec k}^2}   \;, \label{cd41} 
\end{equation}
and 
\begin{equation}
\int \frac{dk_4 \; d^3{\vec k} }{(2\pi)^4} \frac{1}{k^2_4\; {\vec k}^2 + {\vec k}^4 +{2Ng^2\gamma^4} }= \frac{3}{2} \frac{1}{2Ng^2} \;. \label{gp1}
\end{equation}
In this case, both expressions are safe in the IR, while they suffer from UV divergences  which should be handled by a suitable renormalization procedure in the Coulomb gauge. In fact, taking a closer look at the expression \eqref{cd41}, we may write  
\begin{equation}
 \langle {\bar \phi}^{ab}_i \phi^{ab}_i -{\bar \omega}^{ab}_i \omega^{ab}_i \rangle = - Ng^2\gamma^4 (N^2-1)    \frac{1}{\pi^3} \int_0^\infty d\rho \int_0^\infty dr \frac{1}{\rho^2 r^2 + r^4 + 2Ng^2\gamma^4}  \;. \label{j3} 
\end{equation}
where use has been made of three dimensional polar coordinates. Making the change of variables 
\begin{equation}
\rho \rightarrow  (2Ng^2\gamma^4)^{1/4} \;\rho  \;, \qquad r \rightarrow (2Ng^2\gamma^4)^{1/4} \; r \;, \label{ch}
\end{equation}
we get 
\begin{equation}
\langle {\bar \phi}^{ab}_i \phi^{ab}_i -{\bar \omega}^{ab}_i \omega^{ab}_i \rangle = -  (N^2-1){\sqrt{2g^2N} }\;\gamma^2    \frac{1}{2\pi^3} \int_0^\infty d\rho \int_0^\infty dr \frac{1}{\rho^2 r^2 + r^4 + 1} \;.    \label{cd4} 
\end{equation}
To evaluate the integral 
\begin{equation}
I =  \int_0^\infty d\rho \int_0^\infty dr \frac{1}{\rho^2 r^2 + r^4 + 1} \;,  \label{it1}
\end{equation}
we adopt two-dimensional polar coordinates, $(\rho = R \cos\theta, \;r= R \sin\theta)$, obtaining 
\begin{equation}
I =  \int_0^{\sqrt{2} \Lambda}  dR R\int_0^{\frac{\pi}{2}}  d\theta\;  \frac{1}{R^4 \sin^2\theta + 1} \;,  \label{it2}
\end{equation}
where $\Lambda$ is a cutoff. 
From 
\begin{equation} 
\int_0^{\phi} d\theta \; \frac{1}{R^4 \sin^2\theta + 1} = \frac{\arctan(\sqrt{R^4+1} \tan\phi)} {\sqrt{R^4+1}}  \;, \label{it3}
\end{equation}
we finally get 
\begin{equation}
I = \frac{\pi}{2} \int_0^{\sqrt{2} \Lambda}  dR \frac{R}{\sqrt{R^4+1}} = \frac{\pi}{4} {\rm arcsinh}(2\Lambda^2)  \;, \label{it4}
\end{equation}
which diverges logarithmically as $\Lambda \rightarrow \infty$. \\\\We see therefore that, apart from a UV renormalization,  a non-vanishing two-dimensional condensate $\langle {\bar \phi}^{ab}_i \phi^{ab}_i -{\bar \omega}^{ab}_i \omega^{ab}_i \rangle$ might emerge in $4D$. The effect of this condensate on the dynamics of the theory can be taken  into account by introducing the Refined Gribov-Zwanziger action in the Coulomb gauge
\begin{equation}
S_{RGZ} = S_{GZ} + \int d^4x \;\mu^2 \left({\bar \phi}^{ab}_i(x) \phi^{ab}_i(x) -{\bar \omega}^{ab}_i(x) \omega^{ab}_i(x) \right)  \;, \label{rgz}
\end{equation}
where the parameter $\mu^2$ can be obtained order by order by evaluating the effective potential for the operator $ \left({\bar \phi}^{ab}_i(x) \phi^{ab}_i(x) -{\bar \omega}^{ab}_i(x) \omega^{ab}_i(x) \right)$, as done in \cite{Dudal:2011gd} in the case of the Landau gauge. Evaluating now the spatial  two-point gluon correlation function with the refined action \eqref{rgz}, one gets the decoupling solution 
\begin{equation}
\langle A^a_i(k) A^b_j(-k) \rangle_{RGZ} = \frac{\delta^{ab}}{ k^2_4 +{\overrightarrow k}^2 + \frac{2Ng^2\gamma^4}{{\overrightarrow k}^2+\mu^2}} \left( \delta_{ij} -\frac{k_i k_j} {{\overrightarrow k}^2} \right) \;, \label{rgzgluon}
\end{equation}
leading to an equal time transverse form factor of the decoupling type, namely  
\begin{equation}
D^{tr}_{RGZ} ({\overrightarrow k}) = \frac{\sqrt{{\overrightarrow k}^2+\mu^2}}{\sqrt{{\overrightarrow k}^2 \left({\overrightarrow k}^2+\mu^2\right) + 2Ng^2\gamma^4}}  \;. \label{rtrgz}
\end{equation}
Even if being well beyond the aim of the present letter, one might expect that 
the existence of a decoupling type behavior for the transverse gluon propagator 
should entail modifications on the infrared behavior of the ghost, perhaps resulting in a milder behavior of the ghost form factor in the deep infrared region, similarly to what happens in the Landau gauge. It is worth in fact to point out that, 
within the Schwinger-Dyson approach, a decoupling type  solution for the transverse gluon and its consequences on 
the ghost form factor as well as on the Coulomb potential have been already analysed by the authors of \cite{Epple:2007ut}.  We hope to report soon on this important topic, which deserves a more complete and detailed analysis.

\section*{Acknowledgments}
The Conselho Nacional de Desenvolvimento Cient\'{\i}fico e
Tecnol\'{o}gico (CNPq-Brazil), the Faperj, Funda{\c{c}}{\~{a}}o de
Amparo {\`{a}} Pesquisa do Estado do Rio de Janeiro,  the
Coordena{\c{c}}{\~{a}}o de Aperfei{\c{c}}oamento de Pessoal de
N{\'{\i}}vel Superior (CAPES),  are gratefully acknowledged.


\begin{thebibliography}{9}

 

\bibitem{Cucchieri:1996ja} 
  A.~Cucchieri and D.~Zwanziger,
  Phys.\ Rev.\ Lett.\  {\bf 78}, 3814 (1997)
  [hep-th/9607224].
  
\bibitem{Cucchieri:2000hv} 
  A.~Cucchieri and D.~Zwanziger,
  Phys.\ Rev.\ D {\bf 65}, 014002 (2001)
  [hep-th/0008248].

\bibitem{Zwanziger:2002sh} 
  D.~Zwanziger,
  Phys.\ Rev.\ Lett.\  {\bf 90}, 102001 (2003)
  [hep-lat/0209105].
  
\bibitem{Zwanziger:2003de} 
  D.~Zwanziger,
  Phys.\ Rev.\ D {\bf 70}, 094034 (2004)
  [hep-ph/0312254].
  
\bibitem{Greensite:2004ke} 
  J.~Greensite, S.~Olejnik and D.~Zwanziger,
  Phys.\ Rev.\ D {\bf 69}, 074506 (2004)
  [hep-lat/0401003].
  
\bibitem{Zwanziger:2004np} 
  D.~Zwanziger,
  Phys.\ Rev.\ Lett.\  {\bf 94}, 182301 (2005)
  [hep-ph/0407103].
  
\bibitem{Greensite:2004ur} 
  J.~Greensite, S.~Olejnik and D.~Zwanziger,
  JHEP {\bf 0505}, 070 (2005)
  [hep-lat/0407032].
  
\bibitem{Zwanziger:2006sc} 
  D.~Zwanziger,
  Phys.\ Rev.\ D {\bf 76}, 125014 (2007)
  [hep-ph/0610021].
  
\bibitem{Schleifenbaum:2006bq} 
  W.~Schleifenbaum, M.~Leder and H.~Reinhardt,
  Phys.\ Rev.\ D {\bf 73}, 125019 (2006)
  [hep-th/0605115].
  
\bibitem{Watson:2006yq} 
  P.~Watson and H.~Reinhardt,
  Phys.\ Rev.\ D {\bf 75}, 045021 (2007)
  [hep-th/0612114].
  
\bibitem{Epple:2006hv} 
  D.~Epple, H.~Reinhardt and W.~Schleifenbaum,
  Phys.\ Rev.\ D {\bf 75}, 045011 (2007)
  [hep-th/0612241].
  
\bibitem{Watson:2007vc} 
  P.~Watson and H.~Reinhardt,
  Phys.\ Rev.\ D {\bf 77}, 025030 (2008)
  [arXiv:0709.3963 [hep-th]].


\bibitem{Campagnari:2009wj} 
  D.~Campagnari, A.~Weber, H.~Reinhardt, F.~Astorga and W.~Schleifenbaum,
  Nucl.\ Phys.\ B {\bf 842}, 501 (2011)
  [arXiv:0910.4548 [hep-th]].
  
\bibitem{Burgio:2009xp} 
  G.~Burgio, M.~Quandt and H.~Reinhardt,
  Phys.\ Rev.\ D {\bf 81}, 074502 (2010)
  [arXiv:0911.5101 [hep-lat]].
  
\bibitem{Watson:2010cn} 
  P.~Watson and H.~Reinhardt,
  Phys.\ Rev.\ D {\bf 82}, 125010 (2010)
  [arXiv:1007.2583 [hep-th]].
  
  
\bibitem{Leder:2011yc} 
  M.~Leder, H.~Reinhardt, A.~Weber and J.~M.~Pawlowski,
  Phys.\ Rev.\ D {\bf 86}, 107702 (2012)
  [arXiv:1105.0800 [hep-th]].
  
\bibitem{Weber:2011zzd} 
  A.~Weber, M.~Leder, J.~M.~Pawlowski and H.~Reinhardt,
  J.\ Phys.\ Conf.\ Ser.\  {\bf 287}, 012023 (2011)
  [arXiv:1106.3044 [hep-th]].
  
\bibitem{Szczepaniak:2011bs} 
  A.~P.~Szczepaniak and H.~Reinhardt,
  Phys.\ Rev.\ D {\bf 84}, 056011 (2011)
  [arXiv:1106.5528 [hep-ph]].
  
\bibitem{Watson:2011kv} 
  P.~Watson and H.~Reinhardt,
  Phys.\ Rev.\ D {\bf 85}, 025014 (2012)
  [arXiv:1111.6078 [hep-ph]].
  
\bibitem{Campagnari:2011bk} 
  D.~R.~Campagnari and H.~Reinhardt,
  Phys.\ Lett.\ B {\bf 707}, 216 (2012)
  [arXiv:1111.5476 [hep-th]].
  
\bibitem{Watson:2012ht} 
  P.~Watson and H.~Reinhardt,
  Phys.\ Rev.\ D {\bf 86}, 125030 (2012)
  [arXiv:1211.4507 [hep-ph]].
  
\bibitem{Huber:2014isa} 
  M.~Q.~Huber, D.~R.~Campagnari and H.~Reinhardt,
  Phys.\ Rev.\ D {\bf 91}, no. 2, 025014 (2015)
  [arXiv:1410.4766 [hep-ph]].
  
  

\bibitem{Cucchieri:2000gu} 
  A.~Cucchieri and D.~Zwanziger,
  Phys.\ Rev.\ D {\bf 65}, 014001 (2001)
  [hep-lat/0008026].


\bibitem{Cucchieri:2000kw} 
  A.~Cucchieri and D.~Zwanziger,
  Phys.\ Lett.\ B {\bf 524}, 123 (2002)
  [hep-lat/0012024].
  
  
\bibitem{Burgio:2008jr} 
  G.~Burgio, M.~Quandt and H.~Reinhardt,
  Phys.\ Rev.\ Lett.\  {\bf 102}, 032002 (2009)
  [arXiv:0807.3291 [hep-lat]].

\bibitem{Nakagawa:2009zf} 
  Y.~Nakagawa, A.~Voigt, E.-M.~Ilgenfritz, M.~Muller-Preussker, A.~Nakamura, T.~Saito, A.~Sternbeck and H.~Toki,
  Phys.\ Rev.\ D {\bf 79}, 114504 (2009)
  [arXiv:0902.4321 [hep-lat]].
  
  
\bibitem{Burgio:2012bk} 
  G.~Burgio, M.~Quandt and H.~Reinhardt,
  Phys.\ Rev.\ D {\bf 86}, 045029 (2012)
  [arXiv:1205.5674 [hep-lat]].
  
  \bibitem{Gribov:1977wm}  V.~N.~Gribov, Nucl.\ Phys.\ B \textbf{139} (1978) 1.
  

  
\bibitem{Zwanziger:1988jt} 
  D.~Zwanziger,
  Nucl.\ Phys.\ B {\bf 321}, 591 (1989).
  
\bibitem{Zwanziger:1989mf} 
  D.~Zwanziger,
  Nucl.\ Phys.\ B {\bf 323}, 513 (1989).




\bibitem{Capri:2006vv} 
  M.~A.~L.~Capri, D.~Dudal, J.~A.~Gracey, V.~E.~R.~Lemes, R.~F.~Sobreiro, S.~P.~Sorella, R.~Thibes and H.~Verschelde,
  Braz.\ J.\ Phys.\  {\bf 37}, 591 (2007)
  [hep-th/0603167].
  
\bibitem{Capri:2006cz} 
  M.~A.~L.~Capri, V.~E.~R.~Lemes, R.~F.~Sobreiro, S.~P.~Sorella and R.~Thibes,
  Phys.\ Rev.\ D {\bf 74}, 105007 (2006)
  [hep-th/0609212].
  
\bibitem{Alkofer:2000wg} 
  R.~Alkofer and L.~von Smekal,
  Phys.\ Rept.\  {\bf 353}, 281 (2001)
  [hep-ph/0007355].
  
\bibitem{Huber:2009wh} 
  M.~Q.~Huber, K.~Schwenzer and R.~Alkofer,
  Eur.\ Phys.\ J.\ C {\bf 68}, 581 (2010)
  [arXiv:0904.1873 [hep-th]].
  
    
\bibitem{Cornwall:1981zr} 
  J.~M.~Cornwall,
  Phys.\ Rev.\ D {\bf 26}, 1453 (1982).
  
\bibitem{Aguilar:2004sw} 
  A.~C.~Aguilar and A.~A.~Natale,
  JHEP {\bf 0408}, 057 (2004)
  [hep-ph/0408254].
  
\bibitem{Dudal:2007cw} 
  D.~Dudal, S.~P.~Sorella, N.~Vandersickel and H.~Verschelde,
  Phys.\ Rev.\ D {\bf 77}, 071501 (2008)
  [arXiv:0711.4496 [hep-th]].
  
\bibitem{Aguilar:2008xm} 
  A.~C.~Aguilar, D.~Binosi and J.~Papavassiliou,
  Phys.\ Rev.\ D {\bf 78}, 025010 (2008)
  [arXiv:0802.1870 [hep-ph]].

  
\bibitem{Dudal:2008sp} 
  D.~Dudal, J.~A.~Gracey, S.~P.~Sorella, N.~Vandersickel and H.~Verschelde,
  Phys.\ Rev.\ D {\bf 78}, 065047 (2008)
  [arXiv:0806.4348 [hep-th]].
  
\bibitem{Dudal:2008rm} 
  D.~Dudal, J.~A.~Gracey, S.~P.~Sorella, N.~Vandersickel and H.~Verschelde,
  Phys.\ Rev.\ D {\bf 78}, 125012 (2008)
  [arXiv:0808.0893 [hep-th]].
  
\bibitem{Fischer:2008uz} 
  C.~S.~Fischer, A.~Maas and J.~M.~Pawlowski,
  Annals Phys.\  {\bf 324}, 2408 (2009)
  [arXiv:0810.1987 [hep-ph]].
  
  
  
  
\bibitem{Capri:2008ak} 
  M.~A.~L.~Capri, V.~E.~R.~Lemes, R.~F.~Sobreiro, S.~P.~Sorella and R.~Thibes,
  Phys.\ Rev.\ D {\bf 77}, 105023 (2008)
  [arXiv:0801.0566 [hep-th]].
  
  
\bibitem{Capri:2008vk} 
  M.~A.~L.~Capri, A.~J.~Gomez, V.~E.~R.~Lemes, R.~F.~Sobreiro and S.~P.~Sorella,
  Phys.\ Rev.\ D {\bf 79}, 025019 (2009)
  [arXiv:0811.2760 [hep-th]].
  
  
\bibitem{Capri:2010an} 
  M.~A.~L.~Capri, A.~J.~Gomez, M.~S.~Guimaraes, V.~E.~R.~Lemes and S.~P.~Sorella,
  J.\ Phys.\ A {\bf 43}, 245402 (2010)
  [arXiv:1002.1659 [hep-th]].
  
\bibitem{Dudal:2008xd} 
  D.~Dudal, S.~P.~Sorella, N.~Vandersickel and H.~Verschelde,
  Phys.\ Lett.\ B {\bf 680}, 377 (2009)
  [arXiv:0808.3379 [hep-th]].
  
\bibitem{Dudal:2011gd} 
  D.~Dudal, S.~P.~Sorella and N.~Vandersickel,
  Phys.\ Rev.\ D {\bf 84}, 065039 (2011)
  [arXiv:1105.3371 [hep-th]].
  
\bibitem{Epple:2007ut} 
  D.~Epple, H.~Reinhardt, W.~Schleifenbaum and A.~P.~Szczepaniak,
  Phys.\ Rev.\ D {\bf 77}, 085007 (2008)
  [arXiv:0712.3694 [hep-th]].




\end{thebibliography}
\end{document}